\documentstyle[]{mn}

\newcommand{\acknowledgements}[1]{\vspace{7mm} \noindent {\normalsize \bf
Acknowledgements.\,} {\normalsize #1}}

\include{psfig} \psnoisy 

\begin{document}

\title[Contribution of Galaxies to the Background  Hydrogen-Ionizing
Flux] {Contribution of Galaxies to the Background 
Hydrogen-Ionizing Flux}

\author[J. Devriendt et al.]  {Julien E. G. Devriendt,$^1$  Shiv
K. Sethi,$^1$ Bruno Guiderdoni,$^1$ and Biman B. Nath$^2$ \\
$^{1}$ Institut d'Astrophysique de Paris, CNRS, 98bis Boulevard Arago
F--75014 Paris France\\
$^{2}$ Raman Research Institute, Banglore--560080, India}

\maketitle

\begin{abstract}
We  estimate  the evolution of 
the  contribution of galaxies to the  cosmic background flux  
at $912 \rm \AA$ by means  of a 
semi-analytic model of galaxy formation and evolution. 
Such a  modelling has been quite successful in reproducing the optical
properties of  galaxies. We assume hereafter the high--redshift
damped Lyman-$\alpha$ (DLA) systems to be the progenitors of present day 
galaxies, and we design a series  of models which 
are consistent with the evolution of cosmic comoving emissivities in
the  available near infrared (NIR), optical, ultraviolet (UV), 
and far infrared
(FIR) bands along with the evolution of the  
neutral hydrogen content and average metallicity of  
damped Lyman--$\alpha$ systems (DLA).    We  use these  models  to 
compute  the
galactic contribution to the Lyman-limit emissivity and 
background   flux for $0 \simeq z \le
4$. We take into account the absorption of 
Lyman-limit photons by HI and dust in  
the interstellar medium (ISM) of the galaxies.
We find that the background Lyman-limit  flux due to 
galaxies  might dominate  (or be comparable to)  
the contribution from
quasars at
almost all redshifts if the absorption by HI in the ISM is
neglected. The ISM HI absorption results in  a severe diminishing 
of this flux---by almost three
orders of magnitude at high redshifts to between one and two orders at $z \simeq 0$. Though
the resulting galaxy  flux is completely negligible at high
redshifts, it is  comparable to the quasar flux at $z \simeq 0$.

\end{abstract}


\section{Introduction}

The study of the evolution of the ionization  state of
the diffuse intergalactic medium (IGM) is crucial to
understanding  the formation 
of  structures in the universe. An important input in this 
study is the intensity and evolution of the background  hydrogen-ionizing
 flux. The universe is observed to be 
highly ionized at high redshifts (Gunn \& Peterson 1965, 
Giallongo {\it et al. \/} 1994) and it is widely believed that the 
main mechanism of ionization is  photoionization by  photons at and
below
the Lyman-limit. 
 This view is borne out by estimates of  background hydrogen-ionizing
 flux as inferred by the 'proximity effect'  which fixes the 
value of this flux,  $J_{\rm ion } \simeq 10^{-21 \pm 0.5}\, 
\rm erg \, s^{-1} \, cm^{-2} \, Hz^{-1}\, sr^{-1}$ 
(Bajtlik, Duncan  \& Ostriker 1988; Bechtold 1994;  Giallongo {\it et
al. \/} 1996; Cooke {\it et al. \/}  1997) 
for $2 \la z \la 4.5$. It is not clear 
whether the observed quasars can provide sufficient contribution
to   the  flux
implied by this effect at $z \ge 3$ (Miralda-Escud\'e  \& Ostriker 1990; Giroux  \& Shapiro 1996; Haardt \&
Madau 1996; Cooke {\it et al. \/} 1997). In light of this fact, it was
speculated that high-redshift
star-forming galaxies could be the  dominant contributer to this
 background flux (Songaila, Cowie \& Lilly 1990). At smaller 
redshifts, there is a greater uncertainty in the level  of
this  flux (Kulkarni \& Fall 1993; Maloney 1993; Vogel {\it et al. \/}
1995; Reynolds {\it et al. \/} 1995); however, a tight upper limit of $J_{\rm
ion} <
 8 \times 10^{-23} \, \rm erg \, s^{-1} \, cm^{-2} \, Hz^{-1}
\, sr^{-1}$ at $z \simeq 0$
 exists from H$\alpha$ observations (Vogel et al. 1995).
If this flux is entirely due  to  the  background hydrogen-ionizing  photons
then the contribution of quasars falls  short of the 
observed value by nearly an order of magnitude at $z \simeq 0$
 (Madau 1992; Haardt \& Madau 1996).

Recent years have seen tremendous progress in understanding the formation and
evolution of galaxies at moderate and high redshifts. Prominent 
observations which made it possible  are the probing of the evolution
of galaxies with $z \le 1$ by the  Canada-France  Redshift Survey
(CFRS) (Lilly {\it et al. \/} 1995),
the discovery of high-redshift galaxies using the UV drop-out technique 
(Steidel {\it et al. \/} 1996)  in
the Hubble  Deep Field (HDF)
(Madau {\it et al. \/}  1996; Sawicki {\it  et al. \/} 1997;
 Connolly {\it et al. \/}  1997), and a better 
understanding of the evolution of the neutral hydrogen content and 
average metallicity of DLA systems from
 $z \simeq 0$ to $z \simeq 4$ (Lanzetta {\it et al. \/} 1995, Pettini
{\it et al. \/} 1997).
 Several 
theoretical attempts have been made to 
 explain these observations 
(Madau 1996, 1997; Fall {\it et al. \/} 1996; Frenk {\it  et al. \/} 1996;
Guiderdoni {\it et al. \/} 1997b, hereafter GHBM). 
These observations give valuable 
clues  about the history of star formation and the comoving 
emissivity in several  wavebands
in the universe upto $z \simeq 4$ and  have provided a unique 
opportunity to model and 
study the contribution of star-forming galaxies to the background
 hydrogen-ionizing flux (for recent attempts to estimate the 
hydrogen-ionizing flux from star-forming galaxies for  $z \le 1.5$ see
Giallongo {\it et al. \/} (1997); Deharveng {\it  et al. \/} (1997)). 

The CFRS  observations   give the  evolution of cosmic emissivity  at $\lambda
 \simeq  2800 \, \rm \AA$ (in the rest frame of the emitter)
for $z \leq  1$ while the  HDF observations 
 estimate  the emissivity at $\lambda \simeq 1620 \, \rm
 \AA$ for $2.5 \leq z \leq 4$.
 Our aim in this paper is to extrapolate the results from these
important UV wavelengths   to
 calculate  the emissivity  (and background flux) at the Lyman-limit
 wavelength. The hydrogen-ionizing photons are emitted by  massive stars
with lifetimes $\simeq  10^7 \, \rm years$,  so their flux is not 
sensitive to the history of galaxy  evolution, which has much 
larger time scales. Conversely, one can view it as a direct
 measurement of the instantaneous SFR.  
The conversion of emissivity from
the wavelengths probed by HDF and CFRS to $912 \, \rm \AA$ 
requires knowledge of three factors:
\begin{itemize}
\item  the absorption of photons with 
$\lambda \le 912 \, \rm \AA$ within the stellar atmosphere. 
\item the amount of 
dust and the spectral index of dust absorption coefficient in the 
interstellar medium. 
\item the amount of neutral hydrogen in the interstellar
medium. 
\end{itemize}
The first factor is fixed by stellar atmosphere  models and
depends  mainly on the  choice of  IMF which is guided by 
our ability to reproduce observed colours of galaxies.

In extrapolating the amount of dust absorption from the observed 
value at $1620 \, \rm  \AA$ (or $2800 \, \rm
\AA$) to the Lyman-limit wavelength, it is important to know the
amount of dust in the galaxies as accurately as possible, 
because $A_{912}/A_{V} \simeq 5$
 as compared to $A_{1600}/A_{V} \simeq 2.5$. Therefore a small
uncertainty in the dust extinction at UV  wavelengths translates
into a much larger error in the extinction at the Lyman-limit. We can use
FIR data available to fix the amount of dust at $ z \simeq 0$. At
higher redshift, we can fix it by matching the observed  average 
metallicity of DLA
absorbers  (Pettini
{\it et al. \/}  1997) and by reproducing the cosmic infrared background (CIB)
reported by Puget {\it et al. \/} (1996).

The most crucial factor in determining the absorption  of hydrogen-ionizing
photons  in the ISM is the amount of HI and its
distribution in  galaxies. 
We can model this by using the available
 data on DLA absorbers  which give the evolution of the  average HI content
 of these clouds for $0 \leq z \leq 4$ (Lanzetta {\it et al. \/} 1995).
 In our study we have to assume that   DLA
absorbers belong to  the same population of galaxies as the ones that are
 observed in emission in the  CFRS and HDF.
No clear consensus has been reached on this point 
 yet, but recent observations may support
the explanation that the DLA absorbers are either  thick rotating disks
or protogalactic clumps  (Haenelt {\it et al.} 1997;  Prochaska \& Wolfe
1997). Thereafter We assume the HI, dust, and  stars to be homogeneously
 distributed 
in the ISM. This assumption gives the maximum possible HI absorption
 and allows us to obtain a lower limit on the fraction of 
 Lyman-limit photons escaping from galaxies. We shall also briefly discuss the 
implications of the clumpiness of  HI.

We use a semi-analytic model of galaxy formation to 
study the evolution of background hydrogen-ionizing flux
 for $0 \simeq z \la 4$ 
(for details see, e.g.  GHMB). 
 In \S 2, we describe the relevant features this 
 model 
and its  various assumptions.  The basic method of obtaining the
background ionizing flux given the comoving emissivity  and various sources of 
intergalactic absorption is also briefly described in this section.
 Our results are detailed in \S 3.  In \S 4 we discuss  
various uncertainties and possible improvements in our estimates and
summarize our conclusions in \S 5.

\section{Semi-analytic models and Hydrogen-Ionizing flux}

We do not attempt to give a detailed description of
semi--analytic models of galaxy formation and evolution.
Readers interested in details are referred to the recent paper  GHBM
 and references therein. We just
introduce and define the  parameters required for our work.

\subsection{Collapse of halos and cooling}

We consider fluctuations in the matter density distribution at high redshift 
to be spherically symmetric (``Top Hat'' approximation, see eg. 
Peebles 1980)  on galactic scales. As a result of the
violent relaxation process that they undergo after collapsing at redshift
$z_{\it coll}$, such halos virialize.
One can then use the peaks formalism (Bardeen {\it et al.} 1986) to compute 
the mass distribution of such halos, as in
Lacey and  Silk (1991) and Lacey {\it et al. \/} (1993), for
a given cosmological model.\\

The baryonic gas subsequently radiates away its energy, falling deeper
and deeper in the potential well, until it reaches rotational
equilibrium and settles into a disk--like structure. 
Thus, the length scale $r_{\it disk}$, of the cold gas exponential
disk can be inferred from the  
initial radius $r_v$ by conservation of angular momentum 
 $r_{\it disk} \simeq \lambda_J r_v$ where $\lambda_J$ is the
 dimensionless spin parameter (Fall \&  Efstathiou 1980).
Only disks can form in this formalism. 
The formation of elliptical galaxies (and of bulges of spiral
galaxies) has to be 
explained by the merging of disks. Kauffmann {\it et al. \/} (1993) and
Cole {\it et al. \/} (1994) showed that this process can explain the current fraction of
giant ellipticals among bright galaxies. \\ 

\subsection{Star formation}

It is well known that local physical processes governing star
formation are complex and depend on various parameters.  Nevertheless,
it seems that, on galaxy scales, one can phenomenologically define a
mean Star Formation Rate (hereafter SFR) per unit surface density,
which is proportional to the total gas surface density (neutral plus
molecular) divided by the dynamical time scale of the disk (Kennicutt 1989,
1997). Hence we assume that the star formation time scale
$t_\star$ is proportional to the dynamical time scale of the disk:
 $t_\star \equiv \beta(z) t_{\it dyn}$ where $t_{\it dyn} \equiv 2\pi
r_{\it disk} /V_c$. 
In the previous formula $V_c= \left(GM/r_v \right) ^{1/2}$ is the
circular velocity of the dark matter halo of mass $M$, and
$\beta(z)$ is our first free parameter, that we allow to vary with redshift
in order to reproduce the shape of the cosmic comoving emissivity at $1620
\, \rm \AA$ and $2800 \, \rm \AA$ 
 from $z=0$ to $z=4$. The SFR at time $t$ is then computed as follows: 
\begin{equation}
SFR(t) = {M_{\it gas}(t) \over t_\star} .
\label{sfr}
\end{equation}
As shown in GHBM (their fig. 3),
using such a prescription allows one to match 
the average value and width of the
observed distribution of ``Roberts time'' for a $z\simeq 0$
bright disk galaxies sample given by Kennicutt {\it et al.} (1994), 
just by taking $\beta(0) \simeq 100$. 
Furthermore, as pointed out by these authors, the shape of the 
cosmic comoving star formation rate density curve (their fig. 9) suggests 
a strong evolution of $\beta$ with redshift. 
Our model shares the same spirit: 
we consider the star formation history in the universe 
as a sum of contributions from a population of quiescent
star--forming galaxies and a population of starburst galaxies whose
respective proportions are modelled through $\beta(z)$.
We allow  $\beta(z)$ to vary from  $\simeq 0.5$ to $\simeq 200$ 
which respectively corresponds to starburst and quiescent mode of star formation.

\subsection{Stellar feedback}

Stellar feedback is modelled following the original work by Dekel and Silk (1986).
It stems from massive star explosions which expel gas from the
galaxies, preventing further star formation to occur.
Physically, when the thermal energy ejected by supernovae  
becomes greater than the binding energy of the halo,
one expects such winds to be triggered with a certain efficiency. 
One can then compute the mass fraction of stars forming before 
the galactic wind:
\begin{equation}
F_\star \equiv {M_* \over (M_*+M_{\it gas})} 
= (1 + (V_{\it hot}/V_c)^\alpha)^{-1} ,
\end{equation}
where $V_{\it hot}$ and $\alpha$ are obtained through a
fit based on SPH simulations of galaxy formation in which most of the feedback
effect is due to momentum exchange as in Cole {\it et al.} (1994).
But one should bear in mind that there is much uncertainty on these parameters
mostly due to the treatment of supernova remnant interactions
with the interstellar medium. 
Moreover,  as pointed
out by Efstathiou (1992) and Blanchard {\it et al.} (1992), there is
likely to be an overall high--$z$ re--heating of the intergalactic medium 
which could prevent
cooling in halos with circular velocities below $V_{\it c} \sim 20$
 to 50 kms$^{-1}$, and possibly as high as 
$\sim (200)^{1/3} V_{\it c}$ in 
case of adiabatic collapse. To sum up,
the situation is very complicated because we lack a global 
theory of feedback processes. Therefore, we adopt a rather crude but simple 
approach to the problem:
we do not attempt to model the redshift dependence of local or
global feedback in this paper, but we just 
consider the values derived from numerical simulations for a typical ten
percent feedback efficiency of momentum exchange.
They give $\alpha =5$ and $V_{\it hot}=130$ km s$^{-1}$. In the
following, we keep the same value for $\alpha$ and allow our second
free parameter $V_{\it hot}$ to vary between 100 and 130 km s$^{-1}$ .

\subsection{Stellar populations}

We use a coupled model of spectrophotometric and chemical 
evolution in order to compute the age dependence and metallicity
of the gas content, and of the UV to NIR spectra of the stellar 
populations in a self-consistent way. 
The end--product of these models are time dependent synthetic spectra
and gas/metallicity evolution of galaxies.
The model which is used here is based upon a new numerical scheme
(isochrone), as well as on up--graded Geneva stellar tracks and
yields, and will be described in Devriendt, Guiderdoni and  Sadat 
(1998, in preparation). The photometric properties of galaxies
which are our main interest in this study are obtained after taking
into account the intrinsic extinction due
to both the neutral hydrogen absorption below $912 \, \rm \AA$  and the
dust (see following section).

\subsection{Dust and gas absorption}

Some of the energy released by stars in the UV and optical is absorbed
by dust and HI gas (for wavelengths shorter than $912 \, \rm \AA$ ) and 
re--emitted in the IR and submm ranges.   
We would like to emphasize the difficulty of estimating these
absorption processes since
one has to address the complicated and crucial issue of the geometrical
distribution of dust and gas relatively to stars.
In the following, we assume that the gas is
distributed in an exponential disk with mean $\rm HI$ 
column density at time $t$,
\begin{equation}
 \langle N_{\sc\rm HI,ISM}(t)\rangle =M_{\it gas}(t)
/1.4 m_{\sc\rm H} \pi r_g^2,
\label{colden}
\end{equation}
 where
$r_g \equiv f_c \, r_{\it disk}$ is the truncation radius of the gaseous
disk, $f_c$ being our third free parameter.   
The factor 1.4 accounts for the presence of helium.
The mean optical thickness inside $r_g$ is given by:

\begin{eqnarray}
\langle \tau_\lambda (t) \rangle &=& \left( {A_\lambda \over
A_V} \right)_{Z_\odot} \left( {Z_g(t) \over Z_\odot} \right)^s 
\left({ \langle N_{\sc\rm HI,ISM}(t) \rangle \over
2.1~10^{21} {\rm~at~cm^{-2}}} \right) \nonumber \\ 
&+&  \langle
N_{\sc\rm HI,ISM}(t) \rangle \,  \sigma_{\scriptscriptstyle \rm HI}
\left( {\lambda \over 912 {\rm ~\AA}} \right)^3 
\Theta (912 {\rm ~\AA} ),
\label{tau}
\end{eqnarray}
where the first term contains:
\begin{itemize} 
\item the extinction curve for solar
metallicity as measured by Mathis {\it et al.} (1983). 
\item the gas metallicity ($Z_g(t)$) dependence of the extinction curve which
is modelled as in Guiderdoni and  Rocca--Volmerange (1987) and
Franceschini {\it et al.} (1991, 1994),
according to power--law interpolations based on the Solar Neighbourhood and
the Magellanic Clouds, with $s=1.35$ for $\lambda < 2000 \,\rm \AA$ 
 and $s=1.6$ for $\lambda > 2000 \, \rm \AA $.
\end{itemize}
The second term is due to the hydrogen absorption and hence is only used
for wavelengths shorter than $912 \, \rm \AA$ ($\Theta$ being the Heavyside 
function), $\sigma_{\scriptscriptstyle \rm
HI} = 6.3 \times 10^{-18} \, \rm cm^{2}$ is the hydrogen
ionization cross section at the threshold.\\

Light scattering by dust grains is also modelled and the final optical
depth is averaged over inclination angle assuming stars, dust and gas
are homogeneously mixed as in Dwek and  V\'arosi (1996). These
authors extended the
radiative transfer calculation in a spherically symmetric ISM to the
case of an ellipsoid which is more relevant for our study. 
The results are quite similar
to those obtained with a classic ``slab'' geometry where stars,
dust and gas are distributed with the same scale height, and seem  to
satisfy  the estimates of extinction given by
 Andreani and Franceschini (1996). 
As the galaxy is optically thick to Lyman-limit photons, this
homogeneous distribution of stars and dust 
implies that only a tiny fraction ($\simeq 1/ \langle \tau_{912}
\rangle $) of Lyman-limit photons will be able to escape
from the galaxy: the ones emitted by stars located in a thin outer
shell (with column density $\simeq 1/\sigma_{\scriptscriptstyle \rm
HI} \simeq 1.6 \times 10^{17} {\rm~at~cm^{-2}}$)  of the gaseous disk.  
This procedure enables one to compute the bolometric
luminosity absorbed by dust and re--processed in the IR/submm.

\subsection{Dust emission}

Here, we will not give details on how we derive the IR/submm spectrum
(see GHBM), but just briefly outline the steps that lead to the 
emission spectrum.
As mentioned in the previous section, we compute the optical depth 
of the disks, and the amount of bolometric luminosity absorbed 
by dust. The last step consists in the computation of emission spectra of
galaxies in the IR/submm wavelength bands.
This is completed by using a three-component dust model
(Polycyclic Aromatic Hydrocarbons, Very Small Grains and Big Grains),
which is described in D\'esert {\it et al.} (1990).
The method employed is very similar to that developed by Maffei (1994),
and is based upon observational correlations of the IRAS flux ratios
with the total IR luminosity (Devriendt, Guiderdoni \& Sadat
1998, in preparation).      

\subsection{Hydrogen-Ionizing Flux}

Once one has computed the fraction of hydrogen ionizing photons
that effectively escape from the galaxies, one still has to estimate
the optical depth of the inter--galactic medium. This section describes
the prescription we use.
Given the comoving emissivity of hydrogen-ionizing flux 
$\epsilon(\nu,z)$, the specific 
intensity of the background flux $J(\nu_0,z_0)$ (in 
$\rm erg \, s^{-1} \, cm^{-2} \, Hz^{-1}\,sr^{-1}$) 
 is calculated using (Bechtold {\it et al. \/} 1987):  
\begin{eqnarray}
J(\nu_0,z_0) & = &
{c H_0^{-1}\over 4 \pi} \int^\infty_{z_0} dz {(1 + z_0)^3 
\over (1+z)^5 (1+\Omega_0z)^{1/2}}  \nonumber \\
& \times  & \epsilon(\nu,z) 
(1+z)^3 
\exp[-\tau(\nu_0,z_0,z)]
\end{eqnarray}
Here $\tau(\nu_0,z_0,z)$ is the optical depth for a photon  emitted 
at redshift $z$ with frequency $\nu$ and observed at redshift $z_0$ 
with frequency $\nu_0  = \nu (1+z_0)/(1+z)$. 
In this paper we assume that this optical depth is due to HI in 
 Lyman-$\alpha$ 
system clouds encountered along any line of sight. 
 The absorption due to singly ionized helium is
negligible in determining the  hydrogen ionizing flux. Neutral helium
 is always
a sub-dominant species in the Lyman-$\alpha$ clouds in case the background 
flux is dominated
by quasars or star-forming galaxies. Therefore  we neglect the absorption 
due to neutral helium. We also  
neglect the absorption due to diffuse HI in the intergalactic
medium, because there is no  evidence of its presence. 

The line-of-sight average  
optical depth $\tau_{\rm \scriptscriptstyle  Ly}(\nu_0,z_0,z)$
 due to Poisson-distributed Lyman-$\alpha$ clouds can be expressed as
 (Paresce {\it  et al. \/}  1980):
\begin{eqnarray}
\tau_{\rm \scriptscriptstyle Ly}(\nu_0,z_0,z) & = & \int_{z_0}^z \int_0^\infty 
dz \, dN_{\sc\rm HI,IGM}\, {\cal P}(N_{\sc \rm HI,IGM},z) \nonumber \\
& &{} \times \{1-\exp 
\bigl \lbrack  -  N_{\sc\rm HI,IGM}
\sigma_{\sc\rm HI}(\nu) \bigr \rbrack\}. 
\end{eqnarray}
Here  $N_{\sc\rm HI,IGM}$ is the  neutral hydrogen column density of the
Lyman--$\alpha$ clouds. ${\cal P}(N_{\sc \rm HI,IGM},z)$ is the 
 average
 number of clouds with HI column density between $N_{\sc \rm HI,IGM}$ and 
$ N_{\sc \rm HI,IGM}+ dN_{\sc \rm HI,IGM}$ in a redshift range $z$ to 
$z+dz$ along 
any  line of sight. With the existing observations, it is 
not possible to uniquely fix the form of ${\cal P}(N_{\sc \rm HI,IGM},z)$ from
$0 \la z \la 4$. For $z \ge 1.7$ we take model 3 of Giroux \&
Shapiro (1996)
  for Lyman-$\alpha$ clouds, with
 $N_{\sc \rm HI} \le 1.5 \times 10^{17} \, \rm  cm^{-2}$.  For $z \le
 1.7$,  we take 
the redshift evolution given by Bahcall {\it et al. \/} (1993) and normalize the 
number of clouds per unit redshift
at $z = 1.7$ with model 3 of Giroux and Shapiro 1996. This is consistent
with the findings of Bahcall {\it et al. \/} (1993) for clouds with equivalent
width, $W \ge 0.32 \, \rm ~\AA$ ($N_ {\sc \rm HI} = 1.5 \times 10^{14}
 \, \rm   cm^{-2}$). For Lyman-limit systems ($N_ {\sc \rm HI} \ge 1.5
 \times 10^{17} \, \rm  cm^{-2}$),  the results of 
Stengler-Larrea {\it et al. \/} (1995) are taken for
$0 \le z \le 4$. The emissivity 
 $\epsilon(\nu,z)$  is determined from the galaxy evolution models which
fulfill a certain criterion described in the next section. The
contribution of 
ionizing flux  from 'reprocessing' inside Lyman-$\alpha$ clouds is
neglected in our estimates (for details see Haardt \& Madau 1996). 

\section{Results}

\subsection{Parameters}

In the following, we consider the so-called Standard Cold Dark Matter
(SCDM) model with 
$H_0=50$ km s$^{-1}$ Mpc$^{-1}$, $\Omega_0=1$,
$\Lambda=0$, and $\sigma_8=0.67$. We take the baryonic
fraction to be $\Omega_B=0.05$, consistent with primordial
nucleosynthesis.
 Our approach in this
paper is  to work within the framework of one cosmological model
and a single set of values of the  parameters, and to assume that all the 
changes that may occur from altering these choices can be compensated
by appropriately adjusting the free parameters of the semi-analytic
models of galaxy formation. As a matter of fact, it seems  that the 
uncertainties from various
galactic evolution processes  like interactions, mergers, etc. are
more important than the potential influence of the background
cosmology and structure formation models (Heyl {\it et  al.} 1995). 

For simplicity, we take  a Salpeter Initial Mass Function
(hereafter IMF) with index $x=1.35$. 
Stars have masses $0.1 \leq m \leq 120 M_\odot$, and we assume that the mass
fraction  forming  dark objects with masses below 0.1 $M_\odot$
is negligible.
Though we have used a Salpeter IMF throughout this
paper, it is possible that the IMF at high
redshift is top-heavy  and hence produces more UV flux. 
We ran a few models with a top-heavy IMF at high-z and
noticed that  we overestimate the metal production by  a considerable
amount. However, these models may be
allowed if a large fraction of metals are blown away with galactic 
winds  in the IGM. Present 
observations suggest that as much as half the metals at high redshifts
are in small column density Lyman-$\alpha$ clouds (Songaila \& Cowie 1996), so 
it may not be unrealistic to assume that a substantial fraction of 
metals are blown from the galaxies. However, it can enhance 
the Lyman-limit flux by a factor of a few units at high redshifts which
does not affect our results significantly (see the discussion the section 3.3). 

In order to put robust bounds on the hydrogen-ionizing emissivity 
$\epsilon(\nu,z)$  coming
from galaxies, we define two models which are representative of the
uncertainties in the high--z measurement of the $1620 \,  \rm \AA$ cosmic 
emissivity (Madau {\it et al.} 1996, Sawicki {\it et al.} 1997).
Model I gives a fairly good fit of Madau {\it et al.}'s results at
high redshift. Model II matches values determined by Sawicki {\it et al.} (1997).
Both models are completely defined by the three parameters
previously mentioned.
The first free parameter (the SFR efficiency parameter), 
$\beta(z)$ is shown in Fig.~\ref{beta} for both  models.
Its strong evolution with redshift is necessary to reproduce the sharp
decrease of the cosmic
comoving emissivity  observed by the CFRS but the results are not very
sensitive to the shape of the function used provided the peak is
located at the right position $z \simeq 1.5$ and that $ 0.5 \leq 
\beta(z) \leq 200 $. For instance, Fig.~\ref{beta} clearly shows that even
though $\beta(z)$ evolves more strongly in model II (dashed curve)
than in model I, there is no drastic difference in the cosmic emissivity
around the peak in both models. 

The value of the concentration parameter
$f_c$ is quite poorly known but, 
as pointed out in GHBM (their figure  6), the prescription given
above allows one 
to match quite well the measured FIR and mm emission of a flux-limited
sample dominated by  mild starbursts and luminous IR galaxies 
(Franceschini and Andreani 1995) provided one takes
 $f_c \simeq 2.7$.  However, such  observations tend  to pick out
 galaxies with central, dense regions undergoing star formation 
(Sanders \& Mirabel 1996) and  therefore are  biased to give  smaller
values of $f_c$. On the other hand, as argued by Mo {\it et al. \/}
(1997) and Lobo and Guiderdoni (1998) (in preparation),  semi-analytic
models predict too small disk radii in the SCDM as compared to
observations. Hence,  we fix the value of $f_c$ to get the correct 
distribution of DLA (Fig.~\ref{hicol}) which  is based on the 
absorption properties of these clouds. 
$f_c$'s influence concerns mainly the HI distribution~: the larger 
the value of this parameter, the more
extended the HI exponential disks and the fewer the DLA systems. There
is likely to be a redshift dependence of this parameter, simply due to
the fact that mergers have a well--known tendency to concentrate 
the gaseous content of disks. However, we did not try to model that 
effect but rather took an average value of the parameter over
redshift. Averaged values $f_c=7$ for model I and $f_c=5$ for model II,
 allow us, as shown in Fig.~\ref{hicol}, to reproduce quite well
 the column density distribution of DLA at several redshifts. 

Finally, as discussed in the last section, the value of the third 
parameter (the feedback parameter), $V_{\it hot}$, is difficult to
determine. 
However, it has very little influence on the results because it
mainly affects the faint-end slope of the HI mass function, so we just
tuned it in order to get the right amount of HI in DLA systems at 
high--z (Storrie--Lombardi {\it et al.} 1996). Such a requirement yields
values for Model I and II of $120$ and $100 \, \rm  km \, s^{-1}$ respectively.

\subsection{Luminosity Densities}
  Figs.~\ref{madau} and \ref{sawicki} show  results for the 
two models discussed above.
The ``merit'' of a model is judged by its ability to predict 
correctly the evolution of emissivities in various wavebands,
the evolution of $\Omega_{\rm HI}$ and  the average
metallicity of  
DLA  absorbers from $0 \simeq z \le 4$.
 The predicted evolution of the Lyman-limit emissivities---with or 
without absorption by HI in the interstellar medium of  galaxies---
is also shown in  the figures.
As seen in Figs.~\ref{madau} and \ref{sawicki}, the two models
capture fairly well the broad features of the universe upto $z \simeq 4$. 
The only discrepancy between the observations and theoretical
estimates is that these models predict too high a value
of $\Omega_{HI}$ at $z \simeq 0$ (Figs.~3a and~4a). To investigate
this point in detail, we show in Fig.~\ref{hicol} 
 the HI mass function at $z = 0$ (Zwaan et
al. 1997), along with the distribution of HI
column density in DLA absorbers  at three 
 redshifts (Lanzetta {\it et al. \/} 1995).  Although  the agreement at higher redshift is seen to be quite
good, the disagreement at $z = 0$ can be noticed in Fig.~2d. Whereas 
the deviation of theoretical  predictions from the observed HI
mass function at
small HI masses is probably due to our feedback  prescription, 
we also overestimate slightly the HI content for  $10^9 \, 
{\rm M_{\odot}} \le
M_{\rm HI} \le 5 \times 10^{10} \, {\rm M_{\odot}}$.  The 
disagreement at large HI masses is more important because
 much of the hydrogen-ionizing photons are seen to be emitted by
 the largest masses in our
models. Nevertheless, this discrepancy, which would result in an  
overestimate of the HI absorption in the ISM in 
evaluating the hydrogen-ionizing emissivity, is somewhat alleviated  by the
fact that the star formation rate is also higher because of its 
proportionality to the gas mass (Eq.({\ref{sfr})).

The emissivity of Lyman-limit photons depends sensitively on 
dust and HI absorption in the ISM. 
At $z \simeq 0$, the amount of dust in  galaxies is fixed by
matching the energy re-radiated in the FIR and mm bands  with  IRAS
results at $60 \, \rm \mu m$, as shown in Figs.~\ref{madau} and
\ref{sawicki}.  At high
redshifts, there is a  greater uncertainty in the average amount of dust
in galaxies. We use Eq.~(\ref{tau}), which relates the dust optical depth with
the average hydrogen column density $\langle
N_{\sc\rm HI,ISM}(t) \rangle$  and metallicity $Z_g(t)$ of a galaxy, 
to fix the 
dust content in a galaxy to within the observational uncertainties in
the evolution of  $\langle N_{\sc\rm HI,ISM}(t) \rangle$ and $Z_g(t)$. 
The recent detection of the CIB (Puget  {\it et al. \/}
1996; Guiderdoni {\it  et al. \/} 1997a) can also be used to constrain the
history of dust emission 
in the universe.  GHBM  used semi-analytic models of 
galaxy formation to explain the observed CIB. We plot results 
of their models (A) and (E), which  successfully explain Puget et
al.'s
results, in Figs.~(3b) and (4b) along with the prediction of 
our models at $\lambda = 60 \, \rm \mu m$. It is quite clear that
there is a good agreement between the two results. Therefore, we 
believe that our
method allows us to get a fair estimate of  
 dust absorption for $0 \leq z
\leq 4$.  

As is evident from Figs.~\ref{madau} and
\ref{sawicki}, the HI absorption in the ISM is the most
important factor in determining the hydrogen-ionizing emissivity. 
It is customary to compute the emissivity of galaxies
and  multiply by a constant escape fraction to account for this
uncertainty (see e.g, Giallongo {\it et al. \/} 1997). We model this by using 
Eq.~(\ref{tau}) for the optical depth of the neutral hydrogen in a  galaxy. As
discussed above, in a geometry where the gas and stars are
homogeneously
mixed, it just means that a factor $1/\langle \tau_{912} \rangle$ of all the
hydrogen-ionizing photons can escape the galaxy. As $ \langle \tau_{912} \rangle
\gg 1$, the validity of such an assumption depends crucially on the 
relative distribution of stars and neutral hydrogen in the ISM. 
However, it should be pointed out that a homogeneous distribution
gives an   overestimate of the absorption, i.e. an
optically thick but clumpy medium will always result in less
absorption, and therefore the average emissivity we  estimate is a 
secure lower limit on the hydrogen-ionizing emissivity  
from star-forming  galaxies.

\subsection{Galactic Background Lyman-limit Flux}

The evolution of the background flux of Lyman-limit photons
 for various cases discussed above is shown
in Fig.~\ref{ion} along with the available observations at low and
high redshifts. The high redshift data shown in Fig.~\ref{ion} is taken
 from Cooke {\it et al. \/} (1997). Other high redshift proximity effect 
calculations such as the one by Giallongo {\it et al. \/}
 (1996) give $J_{\rm ion} 
\simeq (5 \pm 1) \times 10^{-22}\, \rm erg \, s^{-1} \, cm^{-2} \, Hz^{-1}
\, sr^{-1}$ independent of the redshift for $2 \le z \le 4$.
 The evolution of the quasar contribution is also shown, 
based on the quasar luminosity function derived by Pei (1995). All the 
observational results discussed before  give the
 background hydrogen-ionizing flux, $J_{\rm ion}$,  
as opposed to the Lyman-limit flux we calculate.
They are related as: $J_{\rm ion} = 3/(3+\delta) \times J_{912}$,
 where $\delta$ is the spectral index of the UV background flux
(Miralda-Escud\'e  \& Ostriker  1992). For quasar and 
galaxy  UV backgrounds, $\delta \simeq 1\hbox{--}2$ (see for instance 
 Miralda-Escud\'e  \& Ostriker  1990). We scale all the values of  $J_{\rm
 ion}$ assuming $\delta = 1$, at all redshifts,  in Fig.~\ref{ion}.
 We also show the flux
without dust absorption to gauge the relative importance of  dust
absorption as compared to HI absorption. It should be 
pointed out that the curves without dust absorption are not normalized
 to HDF and CFRS fluxes at other wavebands so they cannot be taken as
 realistic estimates of Lyman-limit flux; their only purpose is to estimate
the relative importance of dust and HI extinction of the Lyman-limit flux.
 As is clearly seen, dust 
absorption decreases the HI flux typically by a factor of a few while
 the HI absorption can diminish it by nearly three orders of
 magnitude at high redshift. 

If the absorption of Lyman-limit photons by HI in the ISM of a
galaxy is neglected, then the background hydrogen-ionizing flux from star-forming
galaxies can be comparable to  or dominate the flux from quasars and
 might make up for the  extra flux which might be
required to explain the proximity effect (Cooke et
al. 1997). However, we believe that it is highly unrealistic to
completely neglect the HI absorption in the ISM. As Fig.~\ref{ion} shows, the
ionizing flux at $z \simeq 3$ decreases by more than  three orders of
magnitude when this absorption
is taken into account. Such a decrease cannot be compensated by
the uncertainties in our analysis and therefore
it seems unlikely that  
star-forming galaxies dominate the hydrogen-ionizing flux at high
redshifts.  The situation is
quite different at smaller redshifts. At $z \simeq 0$, even the lower
limit to galactic flux, (see the lower set of curves in Fig.~\ref{ion}) 
 might be more than  the quasar contribution. As discussed above, 
we overestimate the HI absorption in the 
ISM and so the value of the ionizing flux at $z \simeq 0$
could be higher by a factor of a few over the values shown in Fig.~\ref{ion}.
 Therefore, though the
quasars are likely to dominate the background hydrogen-ionizing flux
at high-$z$, the substantial contribution
 to this flux at smaller redshift is very likely to be due to  
star-forming galaxies.  Our results are in qualitative agreement with
 the recent estimates of Giallongo {\it et al. \/} (1997) and
Deharveng {\it  et
 al. \/} (1997).

The nature of the  ionizing background at high redshifts is also indicated
by the recent observation of He~II at high redshifts (see e.g.,
Davidsen {\it et al. \/} 1996). These observations allow one to 
estimate of the softness parameter, $S_L \equiv J_{912}/J_{228}$. At 
$z \simeq 3$,  $S_L
\simeq 40$ for quasar-dominated background flux  (Haardt \& Madau 1996) but
can be as high as $1000$ for background flux dominated by 
star-forming galaxies (Miralda-Escud\'e \& Ostriker 1990). 
 However, various  uncertainties in  the
modelling of Lyman-$\alpha$ clouds at high-$z$ as well as in the
observations do not allow a firm conclusion
 on the softness of the ionizing  background at high-$z$ (Sethi
\& Nath 1997). We also note that the recent numerical simulations
 of Lyman-$\alpha$ absorbers show  a good match with the observations
if a low value of $J_{912}$ ($\simeq 1 \hbox{--}4 \times  10^{-22}\, \rm erg \, s^{-1} \, cm^{-2} \, Hz^{-1}
\, sr^{-1} $) and a
quasar-like spectrum is adopted (Miralda-Escud\'e {\it et al. \/} 1996, 
Zhang et al. 1996). Therefore, it seems that apart from a few  
 estimates of the proximity effect (Cooke {\it et al. \/} 1997), 
the value of ionizing flux
implied by the observed quasars may suffice to explain other
observations at high-$z$, which is in agreement with our conclusions.

\section{Discussion}

What are the various sources of uncertainties and errors in our
estimates? We have assumed that the stars, dust, and HI are
homogeneously distributed in the galaxy, which, as already discussed,
 is not entirely justified. In this section, we discuss the possible
implications of relaxing this assumption.

Throughout this paper we use Eq.~(\ref{sfr}) and (\ref{tau}) which
give the SFR and optical depth averaged over the entire disk. We
experimented with a few cases in which we considered  the effect of 
{\em local \/} SFR and
 $\tau_\lambda$. We subdivided the disk in shells
of increasing radii and estimated  the SFR and optical depth averaged
over these shells; this enabled us to calculate the amount of
Lyman-limit photons escaping from the disk as a function of the radial
distance. In this case, both the SFR and the optical depth decrease 
towards the outer regions of the galaxy. The aim of this exercise was
to investigate whether the decrease in SFR is less important than the
decrease in the optical depth, and a   relatively larger fraction of 
Lyman-limit photons could escape from outer parts of a
galaxy. However, this does not happen because, for our values of $f_c$, the
disk remains optically thick to Lyman-limit photons  
even at the truncation radius  $r_g$ where the HI column density $N_{\scriptscriptstyle \rm HI,ISM}(r_g) 
= \langle N_{\scriptscriptstyle \rm HI,ISM} \rangle 
/1.6 \times f_c^2 \exp(-1.6 \times f_c)$, which means that even there the
fraction  of
Lyman-limit photons which can  escape remains  $\simeq
1/\tau_{912}$. As both the SFR  and  $\tau_{912}$ are proportional to 
$\langle N_{\sc\rm HI,ISM}\rangle$ (Eqs.~(\ref{colden}) and
(\ref{tau})), the
decrease in the optical depth  is compensated by a similar decrease in
SFR. 
Therefore, the total emergent flux from a galaxy remains   the same
whether   we consider quantities averaged over the
entire disk  or  we sum the contributions of all the shells.

Though it is extremely difficult to accurately model the clumpiness of
the ISM, we give below  qualitative argument to gauge its effect on the HI
ISM absorption.  
Let us assume that all the HI in the ISM is  in the form of
HI clouds with number of clouds per column density 
 $f(N_{\scriptscriptstyle \rm HI, ISM})$ distributed as:
\begin{equation}
f(N_{\scriptscriptstyle \rm HI, ISM}) \equiv
{d {\cal N} \over d N_{\scriptscriptstyle \rm HI, ISM}} = 
A N_{\scriptscriptstyle \rm HI, ISM}^{-\gamma}
\label{distri}
\end{equation}
The optically thin clouds ($N_{\scriptscriptstyle \rm HI, ISM} \le
\sigma_{\scriptscriptstyle \rm HI}^{-1}$) mimic the homogeneous
part of the ISM while the clouds with $N_{\scriptscriptstyle \rm HI, ISM} \ge
\sigma_{\scriptscriptstyle \rm HI}^{-1}$ correspond to the clumpy
ISM. The total average optical depth, along any line of sight,
 due to Poisson-distributed 
clouds with  this distribution is:
\begin{eqnarray}
\langle \tau_{ \scriptscriptstyle \rm ISM}^C \rangle & \equiv & 
\int_{N_{\scriptscriptstyle \rm
HI, ISM}({\rm min})}^{N_{\scriptscriptstyle \rm
HI, ISM}({\rm max})} \left [ 1- \exp (-N_{\scriptscriptstyle \rm HI,
ISM}\sigma_{\scriptscriptstyle \rm HI}) \right ] \nonumber \\
 &\times  &  f(N_{\scriptscriptstyle \rm HI, ISM}) \, \,  d N_{\scriptscriptstyle \rm HI,ISM}.
\label{opdep}
\end{eqnarray}

Our aim is to calculate the absorption from
the clumpy part of the ISM and see  what difference it would  make
if the same amount of matter was distributed homogeneously. To do that, we 
are interested in the ratio of the average optical depth of the clumpy
medium  from Eq.~(\ref{opdep}) and   $\langle \tau_{
  \scriptscriptstyle \rm ISM}^{H}\rangle \equiv \langle
N_{\scriptscriptstyle \rm HI,
ISM}  \rangle  \sigma_{\scriptscriptstyle \rm HI}$, the optical depth
if the HI in the clumpy medium was distributed homogeneously. We
further assume that the ionizing stars are located outside the clouds
in order for the HI absorption to be minimum. For the
distribution given by Eq.~(\ref{distri}),  we get:
\begin{equation}
{ \langle \tau_{ \scriptscriptstyle \rm ISM}^{C}\rangle  \over \langle \tau_{ \scriptscriptstyle \rm ISM}^{H}\rangle} = {(-\gamma + 2) \over (-\gamma + 1)}
{\left [ N_{\scriptscriptstyle \rm
HI, ISM}^{(\gamma +1)}(\rm max) - N_{\scriptscriptstyle \rm
HI, ISM}^{(\gamma +1)}(\rm min) \right ] \over \left [ N_{\scriptscriptstyle \rm
HI, ISM}^{(\gamma +2)}(\rm max) - N_{\scriptscriptstyle \rm
HI, ISM}^{(\gamma +2)}(\rm min) \right ]} {1 \over
\sigma_{\scriptscriptstyle \rm HI}}
\label{clumpy}
\end{equation}
Observations  of HI clouds  in the ISM of our galaxy suggest the clouds  
have a distribution given by Eq.~(\ref{distri}) with $1.6 \le \gamma \le 2.2$ for 
$10^{18} \, {\rm at \, cm^{-2}}  \le N_{\scriptscriptstyle \rm HI, ISM} \le  
10^{22} \, {\rm at \, cm^{-2}}$ (Dickey \& Garwood, 1990),  which corresponds
to the clumpy part of the ISM. 
For these values, $0.01 \le \langle \tau_{ISM}^C \rangle / 
\langle \tau_{ISM}^H \rangle \le 0.03$ 
($\gamma = 2$ is excluded as Eq.~(\ref{clumpy})
is not valid for this value). 
Our  simple analysis suggests that the
average optical depth from the clumpy part of the ISM is negligible as
compared to the homogeneous part; and much of the absorption is caused
by the fraction of HI in the homogeneous medium (or optically thin
clouds) in the ISM. There is  a great amount of uncertainty in
determining the fraction of the  homogeneously distributed HI in the
ISM. If,  for the
purpose of this paper, we roughly assume that the HI is distributed 
equally
between the homogeneous and the clumpy ISM, as might be the 
case for our galaxy,  then our analysis suggests
that the average optical depth of the ISM is halved as compared to the
case when the entire ISM is made up of  homogeneously distributed HI. 

It is known that a significant fraction of young stars form inside
optically thick clouds which suggests that the arguments given above 
underestimate the absorption. On the other hand,   
 most of the ionizing stars form 
 in OB associations. The UV 
photons from these stars  could puncture 
the layer of neutral hydrogen to escape into the galactic halo.
 This fact can be used to explain the existence of HII at large scales
heights (Reynolds 1984) which requires a large
fraction  of the photons to escape in the halo (Dove \& Shull 1994).
However,
it is not clear what fraction of this ionizing flux can escape the
halo of a 
galaxy.  The 
observational status of hydrogen ionizing flux from other galaxies is 
 uncertain.  Recent  observation of 4 low redshift starburst
galaxies (Leitherer {\it et al. \/} 1995) allows one to get only
 upper limits of 57\%, 5.2\%, 11.3\%, and 3.2\% on the
escape fraction (Hurwitz {\it et al. \/} 1997). Other arguments 
based on H$\alpha$ luminosity density of the universe give a  more stringent
upper limit  of $\sim 1 \%$ on the escape fraction (Deharveng et
al. 1997). We note that our simple assumption, without puncturing,
already 
 gives an average escape fraction 
of $0.4 \hbox{--}1 \%$ at small redshifts, which is within the observational
uncertainties. Therefore, this effect seems unlikely to be more than a
 few percents at low redshift. However, at high-$z$, as the star
 formation rate is higher, it might well be that the puncturing effect
allows a larger fraction of ionizing photons to escape.

Throughout this paper we assumed that the damped Lyman-$\alpha$
systems are progenitors of present day galaxies. This
assumption has recently been questioned by Prochaska and Wolfe
(1997). They studied  the kinematical properties of damped Lyman-$\alpha$
systems at high redshift and concluded  that most damped
Lyman-$\alpha$ systems correspond to thick rotating disks with
rotational velocities $\simeq 225 \, \rm km \, s^{-1}$. We compare
these results with the predications of our model. We notice that for
standard CDM model, even though about ten percent of the objects have
rotational velocities
beyond the value of $\simeq 225 \, \rm km \, s^{-1}$, the average 
 rotational velocity  of damped Lyman-$\alpha$
systems is $120 \, \rm km \, s^{-1}$ at $z \simeq 3$,
 which is too small to be
consistent with Prochaska and Wolfe's results. However, this
discrepancy is not serious for various reasons. First, the
relationship between  rotational velocity  and $N_{\sc \rm HI}$
depends sensitively on the cosmology. We tried a structure formation
model with an open universe which has $\Omega = 0.3$. In this case,
the average rotational velocity increases to $160 \, \rm km \,
s^{-1}$ with more than one fourth of the objects with velocities greater
than $\simeq 225 \, \rm km \, s^{-1}$ . This increase is not
unexpected because in open universes
structures of galaxy sizes collapse at higher redshifts as
compared to flat cosmology. Second, the distribution of rotational
velocities for damped Lyman-$\alpha$ systems depends crucially on the 
feedback. An increase in feedback parameter $V_{hot}$ results in an increase in the
average rotational velocity by cutting off contributions from smaller
systems. Such an increase in the feedback might come
from either local uncertainties in feedback (Eq.~(2))
or from a global process like a hot
IGM  with temperature $T \simeq 10^6 \, \rm K$ at high
redshifts and therefore would depend on the thermal history of the
IGM. We also compare the distribution of impact parameters in our
models with the inference of Prochaska and Wolfe and found
 good  agreement  for both cosmogonies. Also,  other recent
analyses  suggest that the evidence of damped
Lyman-$\alpha$ systems being thick rotating disk is not compelling
(Ledoux {\em et al. \/}  1997a,  Ledoux {\em et al. \/} 1997b.).  
Ledoux {\em et al. \/} studied  the kinematical properties of a 
sample with  26 damped Lyman-$\alpha$ systems and  concluded  that whereas
the velocities upto $120 \, \rm  km \, sec^{-1}$ might correspond to
rotations of individual systems, higher velocities probably involve
more than one component (Ledoux {\em et al. \/} 1997a,  Ledoux {\em et
al. \/} 1997b).  A similar conclusion, based on simulations,
 was reached by Haehnelt {\em et al. \/} 1997. Therefore, it seems that 
though the recent studies have thrown light on the nature of damped
Lyman-$\alpha$ systems at high-$z$, it is  still too early to draw firm
conclusions. However, should it turn out  that damped Lyman-$\alpha$
systems at high $z$ correspond to merging protogalactic HI  clumps rather
than rotating disk, our conclusions about the HI absorption inside the
ISM of a high-$z$  galaxy  and consequently the Hydrogen-ionizing flux
  will be significantly weakened in the sense that it will be more
  difficult to support the view that sites where a high HI column
  density is detected also correspond to star forming regions.

\section{Conclusion}

We estimated the contribution of star-forming galaxies
to the background Lyman-limit flux taking  into account the HI and dust
absorption  in the ISM of individual galaxies in a self-consistent way
with the cosmic star-formation  history. We assumed  that DLA systems 
correspond to star-forming regions at high-redshift. We conclude that while
 star-forming galaxies are unlikely to dominate the
 background hydrogen-ionizing flux at high redshift, 
they are most likely to do so in the present universe. The current
uncertainties of modelling do not allow us to calculate the redshift
of cross-over from quasar-dominated to galaxy-dominated background
flux. As already discussed, a good discriminator between these  two
sources  of ionizing-background  is
the softness of their spectra. Future high-resolution observations of
metals (e.g. Carbon) in
their various ionization states in  the low-column density Lyman-$\alpha$
clouds  for $0 \simeq z  \le 4$ might enable one to 
find  the transition from quasar-dominated  to galaxy-dominated
 background flux.

\acknowledgements{We are pleased to thank Patrick Petitjean,
  R. Srianand and the anonymous referee for pertinent comments 
on the kinematics of DLAs.}

\clearpage
\begin{figure*} 
\centerline{\psfig{figure=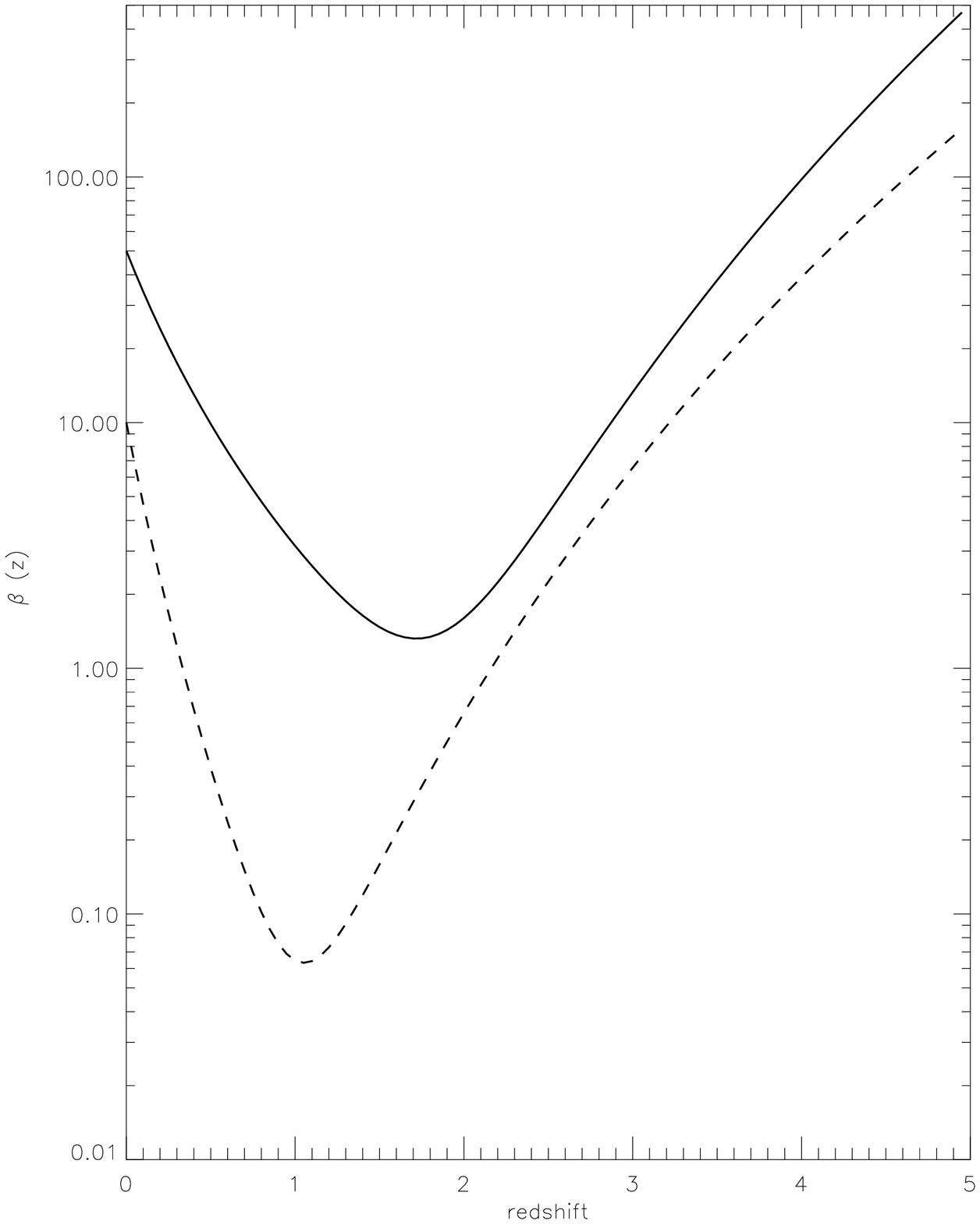,width=0.6
\textwidth}}
\caption{Efficiency parameter evolution with redshift. The {\em thick
  solid \/} and {\em dashed \/} lines show the values used for model I and II
  respectively. Analytical expressions for $\beta(z)$ are 
$\beta(z)=50(1+10^{-6} (1+z_{\it coll})^{13})/(1+z_{\it coll})^4$
  (model I) and $\beta(z)=10(1+10^{-5}(1+z_{\it coll})^{16})
/(1+z_{\it coll})^8$ (model II).} 

\label{beta}
\end{figure*}
\clearpage
\begin{figure*} 
\centerline{\psfig{figure=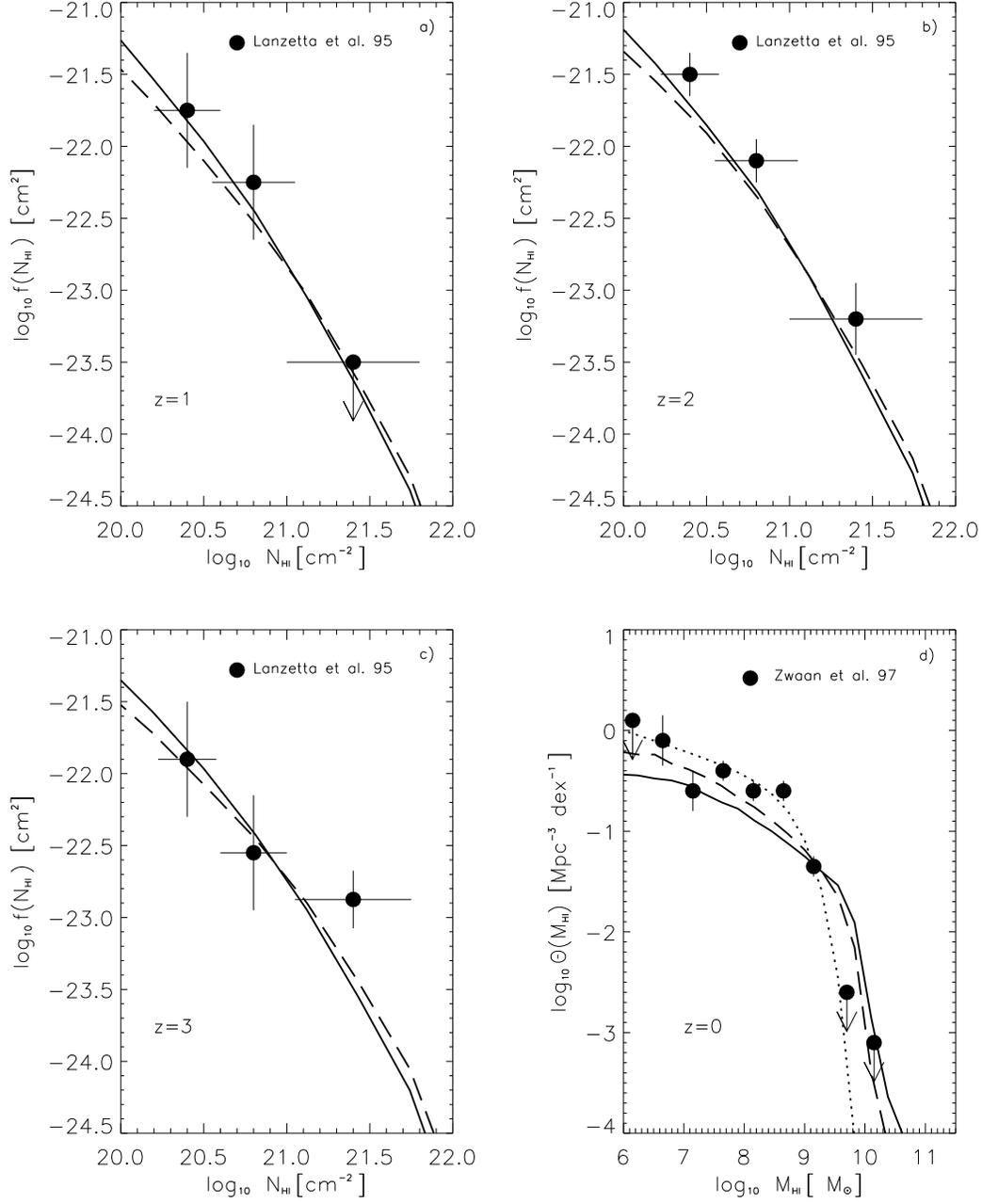,width=0.8\textwidth}}
\caption{Fig. 2a: Distribution of damped Lyman--$\alpha$ absorbers as a
function of their HI column density at $z \simeq 1$. {\em Thick solid \/} and
{\em long-dashed \/} lines are respectively the predictions of models I and II.
Fig. 2b: Same as panel a) for $z \simeq 2$ .
Fig. 2c: Same as panel a) for $z \simeq 3$.
Fig. 2d: HI mass function of galaxies at $z \simeq 0$. {\em Thick
solid \/}, {\em long dashed \/}, and  {\em dotted \/} curves are respectively models I and II
predictions and a Schechter fit to the data given by Zwaan et
al. 97.}
\label{hicol}
\end{figure*}

\clearpage
\begin{figure*} 
\centerline{\psfig{figure=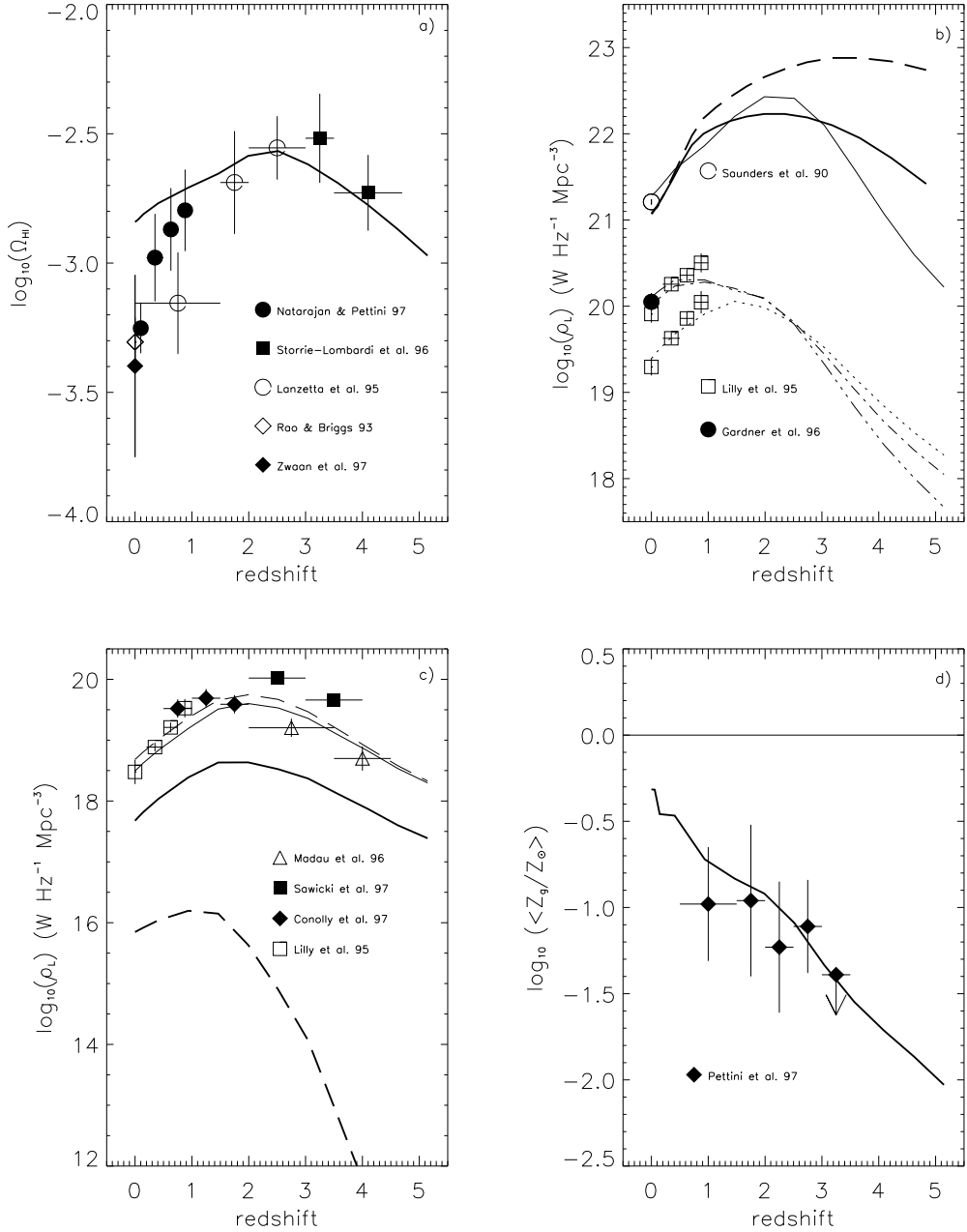,width=0.8\textwidth}}
\caption{Fig. 3a:  Evolution of the cold gas density parameter in 
DLA absorbers. The fit from Model I is indicated by the {\em thick 
solid \/} line.
Fig. 3b: Rest--frame comoving luminosity density from optical to FIR. 
The {\em thin dotted \/}, {\em  dot-dashed \/}, {\em triple-dot-dashed
\/} and {\em solid \/} 
lines respectively stand for emissivities at
4400 ${\rm \, \AA}$, 10000 ${\rm \, \AA}$ 22000 ${\rm \, \AA}$ and 
60 $\mu \rm m$ given by model I.
Open squares: local and Canada--France Redshift Survey (Lilly et
  al. 96). Open circle: 60 $\mu$m local density
corresponding to one third of the bolometric light radiated in the IR
(Saunders et al. 90). Solid circle : 2.2 $\mu$m local
luminosity density (Gardner et al. 96).
The {\em thick solid \/}  and {\em dashed \/}  lines are respectively models (A) and (E) 
from GHBM that give reasonable fits of the CIB.
Fig. 3c : Rest--frame comoving luminosity density from far--UV to UV. 
The {\em thin solid \/}  and {\em long-dashed \/} lines,
 represent emissivities 
at 1620 ${\rm \, \AA}$ and 2800 ${\rm \, \AA}$  given by model I. 
The {\em thick solid \/}  and {\em long-dashed  \/} 
lines are predictions of the
same model for luminosity at 912 ${\rm \, \AA}$ respectively without and with 
HI absorption, but with dust absorption. 
Solid diamonds: photometric redshifts in the Hubble Deep Field (Connolly
 et al. 97) taking into account NIR data. Solid squares: other
 estimates of photometric
redshifts in the HDF (Sawicki et al. 97). Open triangles:
HDF with redshifts from Lyman--continuum drop--outs (Madau et
  al. 96).
Fig. 3d : Evolution of the mean metallicity in damped
Lyman--$\alpha$ absorbers. Prediction from Model I is indicated by 
the {\em thick solid \/} line. Since the chemical
evolution model is a closed--box one, the metallicity of the
systems is certainly overestimated if a fraction of the metals
is ejected in the IGM.}
\label{madau}
\end{figure*}
\clearpage
\begin{figure*} 
\centerline{\psfig{figure=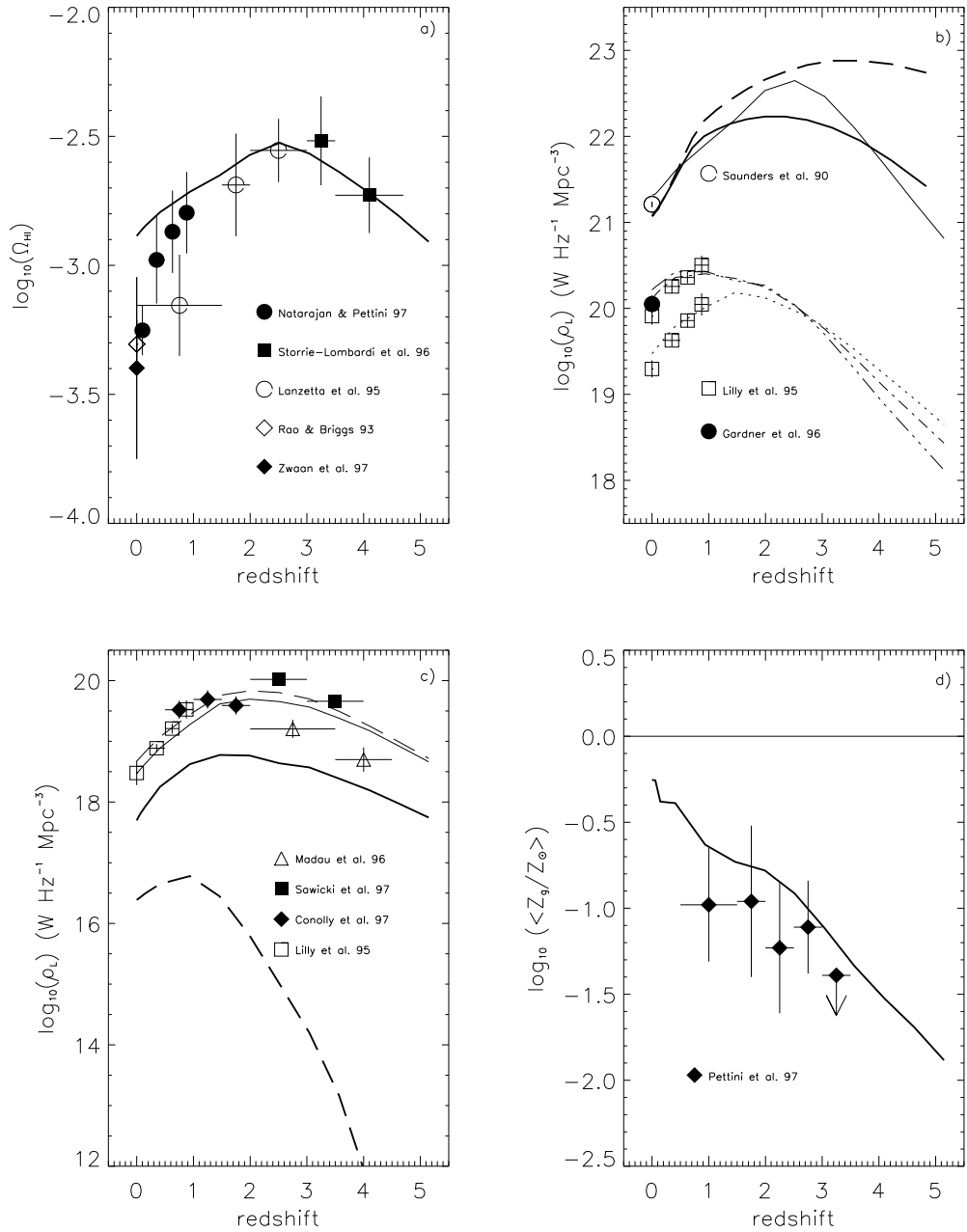,width=0.8
\textwidth}}
\caption{Same as Fig. 3 for model II.}
\label{sawicki}
\end{figure*}

\clearpage
\begin{figure*} 
\centerline{\psfig{figure=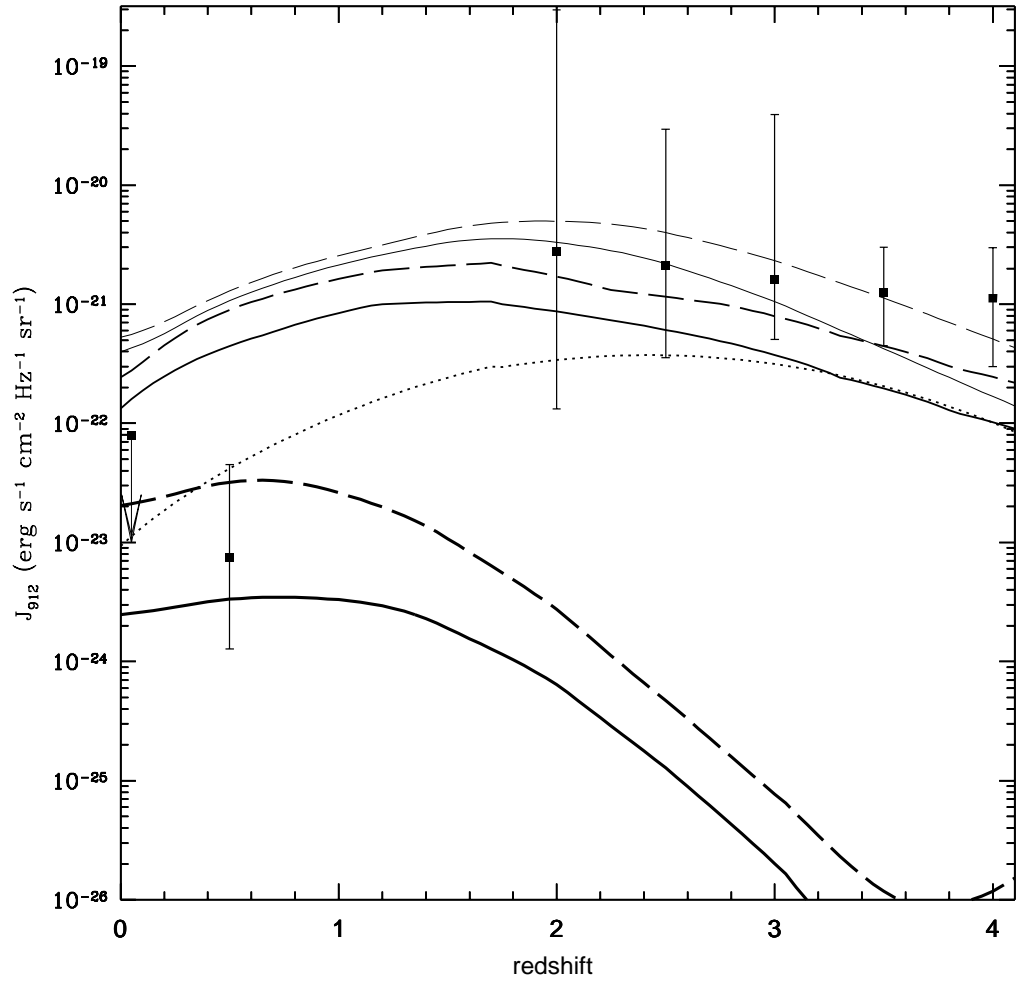,width=0.8
\textwidth}}
\caption{ The evolution of Lyman-limit flux  is shown for the 
various cases discussed in the text. The  
{\em solid\/} lines give the predictions of   Model I. The lines of
increasing thickness correspond respectively to models without dust
and HI absorption, with dust but without HI absorption, and  with both dust
and HI absorption. 
   The  {\em dashed\/}
lines are the corresponding predictions of Model II.  The {\em dotted \/} 
 line  shows  the  
evolution of  background Lyman-limit flux from quasars. 
The upper
limit at $z \simeq 0$ is from Vogel et al. (1995), the data point at 
$z \simeq 0.5$ is from Kulkarni and Fall (1993), the high redshift
data points are from Cooke et al. (1997).}
\label{ion}
\end{figure*}

\end{document}